\documentclass[graphicx,amsmath,11pt,aps,prx,twocolumn,reprint,showpacs,showkeys]{revtex4-1}
\usepackage{graphicx}
\usepackage{hyphenat}
\usepackage{color}

\newcommand{\figs}[1]{Figure \ref{#1}}
\newcommand{\figss}[1]{Figures \ref{#1}}
\newcommand{\fig}[1]{Fig \ref{#1}}

\begin{document} 


\title{Giant anomalous self-steepening in photonic crystal waveguides}

\author{Chad Husko$^{1}$ and Pierre Colman$^{2,3}$}
\affiliation{
$^1$Centre for Ultrahigh bandwidth Devices for Optical Systems (CUDOS), Institute of Photonics and Optical Science (IPOS), School of Physics, University of Sydney, NSW 2006, Australia\\
$^2$ DTU Fotonik, {\o}rsted plads, Kgs. Lyngby, Denmark\\
$^3$ Institut d'Electronique Fontamentale (IEF), Universit\'{e} Paris-Sud, Orsay, France\\
Corresponding author: pierre.colman@espci.org
}

\date{\today}

\begin{abstract}
Self-steepening of optical pulses arises due the dispersive contribution of the $\chi^{(3)}(\omega)$ Kerr nonlinearity. In typical structures this response is on the order of a few femtoseconds with a fixed frequency response. In contrast, the effective $\chi^{(3)}$ Kerr nonlinearity in photonic crystal waveguides (PhCWGs) is largely determined by the geometrical parameters of the structure and is consequently tunable over a wide range.  Here we show self-steepening based on group-velocity (group-index) modulation for the first time, giving rise to a new physical mechanism for generating this effect. Further, we demonstrate that periodic media such as PhCWGS can exhibit self-steepening coefficients two orders of magnitude larger than typical systems. At these magnitudes the self-steepening strongly affects the nonlinear pulse dynamics even for picosecond pulses. Due to interaction with additional higher-order nonlinearities in the semiconductor materials under consideration, we employ a generalized nonlinear Schr\"odinger equation numerical model to describe the impact of self-steepening on the temporal and spectral properties of the optical pulses in practical systems, including new figures of merit. These results provide a theoretical description for recent experimental results presented in [\textit{Scientific Reports 3, 1100 (2013)} and \textit{Phys. Rev. A 87, 041802 (2013)}]. More generally, these observations apply to all periodic media due to the rapid group-velocity variation characteristic of these structures.
\end{abstract}



\maketitle


\section{Introduction}

A recent trend in nonlinear optics is the development and design of waveguide systems with tunable nonlinearities. In addition to a broad tuning range, these systems are characterized by the ability to separate the contributions of the material constituents from the device geometry. This is in contrast to early approaches in both glass \cite{cohen1979tailoring} and semiconductors \cite{aitchison1997nonlinear} which required a change in material composition to modify the waveguide properties. The role of geometry drastically changed with the advent of micro-structured fibers and the demonstration that fabrication parameters could be the dominant contribution to the dispersion \cite{knight1996all}. More recently, it has been shown that gas-filled hollow-core fibers can activate or suppress nonlinearities such as the Raman effect \cite{russell2014hollow}. In parallel, rapid advances in integrated semiconductor devices have pushed the forefront of optical science by reducing nonlinear thresholds to sub-femtojoule energy levels \cite{nozaki2010sub} while simultaneously incorporating dispersion control \cite{Colman:12}. Among nanostructures, photonic crystal waveguides (PhCWGs) are of extreme interest due to the link between geometric fabrication and direct modulation of the electric field, giving rise to new physical phenomena such as slow-light and enhanced nonlinearity. 

Slow-light refers to light propagating  at a reduced group-velocity in the medium. Interest in this unique property has inspired a large body of research investigating the linear and nonlinear properties of slow-light in two-dimensional (2D) PhCWGs over the past decade \cite{baba2008slow,PhysRevE.64.056604}. Recall the group index $n_g$ is related to the waveguide dispersion relation $\omega(k)$ and the group-velocity $v_g$: $n_g=\frac{c}{v_g}=c\frac{\partial\omega}{\partial k}$, with frequency $\omega$, wavevector $k$, and the speed of light in vacuum $c$. Of particular significance, it was shown that optical $\chi^{(3)}$ effects such as the Kerr nonlinearity scale with the group index squared in the presence of slow-light \cite{PhysRevE.64.056604}. Briefly, one factor of $n_g$ arises from a larger electric field for a given power (nonlinear enhancement), with the second from longer effective optical path length (linear enhancement).  We note that the slow-light-enhancement described here is derived from the \textit{structure}. In contrast, \textit{material} slow light from atomic resonances does not exhibit this enhancement \cite{boyd2011material}. In PhCWGs we write the effective nonlinear Kerr parameter as $\gamma_{eff} =\gamma \left(\frac{n_g}{n_o}\right)^2=\frac{\omega}{c} \frac{n_2}{A_{eff}}\left(\frac{n_g}{n_o}\right)^2$, with the bulk Kerr coefficient $n_2$, modal area $A_{eff}$, and linear refractive index $n_o$. While the $\gamma$ term is well described in the literature, research into the slow-light enhancement contribution $\frac{n_g}{n_o}$ in 2D PhCWGs required significant advances in nanofabrication techniques which were only mastered the past few years.

A 2D PhCWG consists of a periodic array of low-index dielectric embedded in a high-index material. A common experimental configuration which we consider here consists of a hexagonal pattern of air holes etched in an air-suspended semiconductor slab. Importantly, the dispersion of these 2D PhCWGs is highly tunable due to selected geometric modifications of the periodic lattice known as \textit{dispersion engineering} \cite{Colman:12,Li:08}. The precise modulation of the waveguide group index enables exquisite control over the dispersion and therefore the nonlinear properties of the medium. Experimental reports of slow-light enhanced nonlinear Kerr effects in 2D PhCWGs include demonstrations of solitons, third-harmonic generation, and four-wave mixing, amongst others \cite{NatPhot_Colman10,corcoran2009green,mcmillan2010FWM}. Despite this strong interest in slow-light, the dispersion of the Kerr $\chi^{(3)}(\omega)$ nonlinearity, or \textit{self-steepening} (SS) term $\tau_{NL}=\frac{1}{\gamma_{eff}}\partial_\omega \gamma_{eff}$, has received surprisingly little attention in these systems.

The earliest investigations of SS in waveguide systems were carried out in glass fiber with these studies emphasizing the spectral re-shaping properties \cite{fork1983femtosecond,fiberSS} or the formation of optical shock fronts \cite{anderson1983}. Later it was shown SS is essential for extending the validity of the Generalized Nonlinear Schr\"odinger Equation (GNLSE) envelope approximation down to the single-cycle regime \cite{brabec1997nonlinear}, and for explaining broadband supercontinuum generation \cite{gaeta2000catastrophic}. In these systems the SS term is determined almost exclusively by the wave angular frequency ${\omega}$ and therefore exhibits a fixed response of about a femtosecond for optical frequencies. A more recent numerical investigation showed the wavelength dependence of the nonlinear Kerr effect in silicon channel waveguides leads to values up to tens of femtoseconds near mode cutoff \cite{siliconSS}. An alternative approach for generating SS using cascaded $\chi^{(2)}$ media as an \textit{effective} tunable $\chi^{(3)}$ was shown with similar strength to traditional media \cite{moses2006}. To date, self-steepening in tunable $\chi^{(3)}$ media has not been explicitly experimentally demonstrated.

In this article, we investigate self-steepening in 2D photonic crystal waveguides. Importantly we show the large variation in waveguide group index $n_g$ leads to a new physical mechanism for generating self-steepening with a characteristic time scale $\tau_{NL}$ on the order of hundreds of femtoseconds, two orders of magnitude larger than in non-periodic waveguide systems \cite{fiberSS,siliconSS}. We derive an analytic formulation and describe the origin of this effect. Further, we describe structures in which the values of $\tau_{NL}$ are \textit{anomalous} (negative), hence leading to notably different physical effects than previously known $\chi^{(3)}$ systems. The broad tuning range of $\tau_{NL}$ enabled by dispersion-engineering make PhCWGs an ideal system for further studies of SS. While the magnitude of $\tau_{NL}$ is quite large, the presence of other effects such as group velocity dispersion (GVD,~$\beta_2=\frac{1}{c}\frac{\partial n_g}{\partial\omega}$), and higher order nonlinearities such as multi-photon absorption or free-carrier effects can disrupt the ideal dynamics. We consequently describe the experimental situations in which SS is expected to contribute significantly in the semiconductor system under consideration using a numerical model. This analysis supports recent experimental results showing pulse temporal advance in PhCWGs \cite{Husko_SciRep13,Raineri_PRB13}. Though the Raman-effect is narrowband in semiconductors and negligible here, this analysis could be extended to include periodic glass media where Raman and Brillouin effects must be considered \cite{freeman2005chalco}. These results provide the first theoretical description of giant and tunable self-steepening in nanoscale optical waveguides.

More generally, this investigation applies to all periodic media (1D, 2D, 3D) where a dispersive $\chi^{(3)}(\omega)$ arises due to strong group-index modulation near the band-edge. Given the importance of the dispersive nonlinearity $\tau_{NL}$ in explaining supercontinuum broadening in fibers, we expect the new terms elucidated here to be critical for accurately describing this phenomenon in photonic crystals. 


\section{Self-steepening in PhCWGs}

\begin{figure}[*htb]
\centerline{\includegraphics[width=9cm]{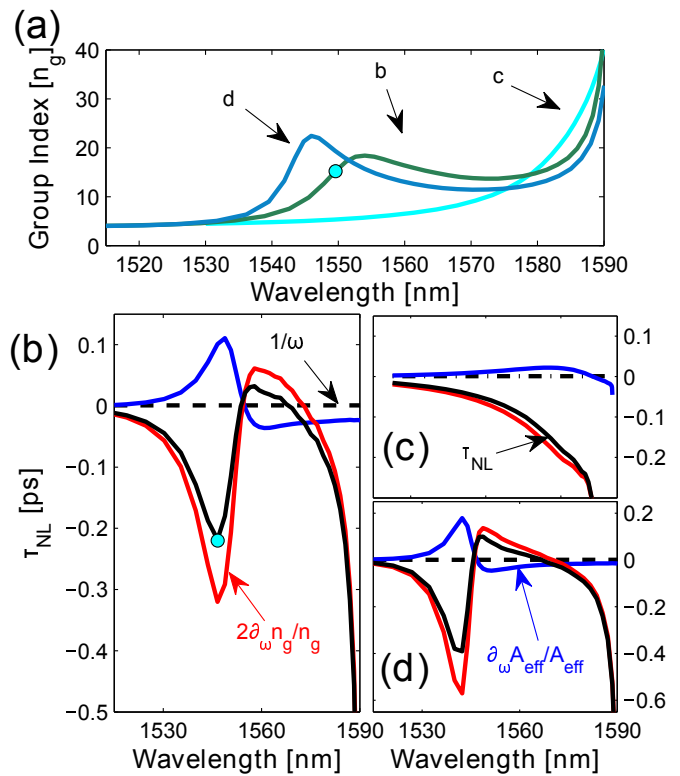}}
\caption{Group index curves and self-steepening parameters of PhC waveguides. (a) We show three typical waveguides: (b,green) a \textit{dispersion-engineered} structure with a quasi flat plateau, (c,cyan) a W1 line-defect waveguide, and (d,blue) and a \textit{dispersion-engineered} waveguide with a peak. (b)-(d) Show the self-steepening parameter $\tau_{NL}=\gamma_1/\gamma_{eff}$ (black line) as a function of wavelength for the three structures. The different contributions to SS are shown: $1/\omega$(black-dashed), $2\partial_\omega n_{g}/n_g$(red) and $\partial_\omega A_{eff}/A_{eff}$(blue).}
\label{Fig:SS_PhCWG} 
\end{figure}

Though near-arbitrary dispersion profiles are possible in periodic media \cite{Colman:12,Li:08}, here we focus on three specific experimentally demonstrated structures for clarity. \figs{Fig:SS_PhCWG} shows three group index curves for PhCWGs with different dispersion relations: (i) a standard line defect waveguide of one missing row of holes in a hexagonal lattice (W1), (ii) a dispersion engineered waveguide\cite{Colman:12} exhibiting a plateau, and (iii) a dispersion with a pronounced group index peak. The dot indicates a point of interest we investigate in this work.

If we assume the Kerr nonlinearity evolves linearly then the impact of dispersive nonlinearity on pulse propagation dynamics is modeled by adding a first order Kerr correction term to the GNLSE \cite{brabec1997nonlinear,siliconSS}:

\begin{equation}
\partial_z A+\frac{i}{2}\beta_2\partial_{tt}A = i \gamma_{eff}(1 + i \tau_{NL} \partial_t) |A|^2A.
\label{eq:GNLSE}
\end{equation}

\noindent Here \textit{t} is the relative time in the reference frame of the pulse with $A(z,t)~=~\sqrt{P_o(z,t)}e^{i\phi(z,t)}$ the electric-field envelope with power $P_o$ and phase $\phi$. The last term $\tau_{NL}$ is referred to the \textit{self-steepening} or \textit{shock term} and expressed as:

\begin{equation}
\tau_{NL}=\frac{\gamma_1}{\gamma_{eff}}=\frac{1}{\omega}+\frac{1}{n_{2}}\frac{\partial n_{2}}{\partial\omega}-\frac{1}{A_{eff}}\frac{\partial A_{eff}}{\partial\omega}+\frac{2}{n_g}\frac{\partial n_g}{\partial\omega},
\label{eq:dOmega}
\end{equation}

\noindent where we used the relationship $\gamma_1~=~\partial_\omega\gamma_{eff}$. The two first terms exist in bulk material and contribute up to a few femtoseconds. In practice the Kerr dispersion is often neglected. These are the \textit{traditional} self-steepening terms known in unstructured waveguides and do not play a role here with these small magnitudes \cite{fiberSS}. The third term encompassing effective modal areas $A_{eff}$ was considered in fibers \cite{fiberModeArea}, and theoretically in silicon channel waveguides \cite{siliconSS} with surprisingly little attention in PCFs \cite{Dudley:06}. Further, in those earlier works the area term contributed a few percent whereas here the slow-light modes exhibit rapid variation in spatial profile with $\omega$ and this term contributes approximately $\approx$~25~\% to the total $\tau_{NL}$. 

The key physical insight of this work is the realization that the final term due to the dispersive group-index $n_g$ gives rise to a previously unknown mechanism for controlling self-steepening. In contrast to earlier observations, in our photonic crystal waveguide the magnitude of $\tau_{NL}$ is set by a combination of the group-index $n_g$, GVD ($\beta_2$) and effective modal area $A_{eff}$. As these three parameters depend strongly on the geometry, we clearly see the advantage of using nanostructures to study the dispersive nonlinearity. We now examine this effect in detail.

\figss{Fig:SS_PhCWG}(b)-(d) show the evolution of the ratio $\tau_{NL}$ for the waveguides in \fig{Fig:SS_PhCWG}(a). The contribution of the different terms composing Eqn. \ref{eq:dOmega} are also represented. We note three major observations unique to the PhCWG system. First, $\tau_{NL}$ is two orders of magnitude larger than unstructured waveguides (black dashed line) where $\tau_{NL}=\frac{1}{\omega}\approx$~1~fs.  Notice the $\frac{1}{\omega}$ contribution appears to be near-zero and completely flat compared to the PhCWG contributions on this scale. Second, the sign of $\tau_{NL}$ is negative. Consequently the spectral and temporal properties of the nonlinear waves behave in an opposite or \textit{anomalous} manner as we will examine. Third, the dominant contribution arises from the dispersion term (red) with a modification in the opposite direction due to $A_{eff}$ (blue). If we ignore the area contribution, we approximate $\tau_{NL} \approx \frac{2}{n_g}\frac{\partial n_g}{\partial\omega}=\frac{2c\beta_2}{n_g}$.

Examining the characteristics of the waveguides individually, we find different trends for each. Regarding the W1 waveguide \fig{Fig:SS_PhCWG}(c), $\tau_{NL}$ steadily decreases approaching the mode cutoff (increasing wavelength). A value of about -200 fs is obtained close the the band edge, however in a region where propagation loss is large in this type of PhCWG \cite{OFaolain2010loss}. On the contrary for the dispersion-engineered waveguides the values are mostly negative with a small positive region for the structures presented. Notice in this case values as large as -200 fs (\fig{Fig:SS_PhCWG}(b)) and -400 fs (\fig{Fig:SS_PhCWG}(d)) are reached away from the band edge where propagation losses are reduced \cite{Patterson_PRB2010,Mann_OL13}. The lower linear loss of the latter structures has implications for practical observation of these effects.

\section{Temporal and spectral properties due to anomalous self-steepening in PhCWGs}

\par We now describe the physical implications of the self-steepening term on nonlinear wave propagation in PhCWGs. For that purpose we consider typical parameters found in recent nonlinear experiments \cite{Husko_SciRep13,Colman2012,NatPhot_Colman10,Raineri_PRB13,combrie2009}. We take $\gamma_{eff}$~=~1600~(W.m)$^{-1}$ ($n_g$=15, $n_o$=3.17 for GaInP), an anomalous dispersion of $\beta_2$~=~-7.7 ps$^2$/mm, $n_2$~=~6$\times10^{-18}$m$^2$/W and modal area $A_{eff}$~=~0.34 $\mu$m$^2$ \cite{Colman:12}. The dispersive nonlinearity is $\tau_{NL}$=~-220 fs as detailed above. The input pulses are $T_{FWHM}$~=~2.3 ps (full-width at half-maximum of a hyperbolic secant, $T_{FWHM}=1.76~T_{o}$) with $P_0$~=~3~-~10~W~(6 - 20 pJ/pulse). The dispersion length $L_D=\frac{T_o^2}{\beta_2}$ is computed as 220~$\mu$m. Importantly, throughout this work we purposely maintain small soliton numbers ($N<$~2) so as to avoid more complicated soliton dynamics modulating the peak intensity and pulse duration \cite{Husko_SciRep13}. 

At this point we are focusing on the basic physical effects resulting from the unique photonic crystal dispersion in the `ideal' system. In the next section we will introduce the full 
effects present in typical semiconductor waveguides and describe how these results are modified. While based on actual experimental structures, note that the conditions may not be optimal for emphasizing the self-steepening effect and we invite the community to explore the parameter space further.

The normalized self-steepening parameter $s$ is:
\begin{equation}
s = \frac{\tau_{NL}}{T_0}.
\label{eq:shockParameter}
\end{equation}

\noindent For our parameters $s$=~-0.1, more than five times larger than non-periodic waveguides \cite{fiberSS}. This is even more remarkable when one considers the pulses are 2.3 ps long compared to the sub-100 fs pulses required in unstructured media where $s$~=~$\frac{1}{\omega T_0}$. The large value of $s$ requires much shorter length scales to observe the associated effects of self-steepening. A pulse experiencing self-steepening will eventually develop a shock front after propagating a \textit{shock length}\cite{anderson1983} of about:
\begin{equation}
z_s = 0.43 \frac{L_{NL}}{|s|}, 
\label{eq:shockLength}
\end{equation}

\noindent
where $L_{NL}~=~(\gamma_{eff} P_o)^{-1}$ is the nonlinear effective length, and the numerical constant depends on the actual pulse shape, here a hyperbolic secant \cite{anderson1983}. Typical shock lengths in our structures are $z_s\approx$ 350 $\mu$m for a 2.3 ps pulse with peak power of 8 W. As this article is focused on self-steepening, we do not describe the physics of optical shock waves in detail here. Nonetheless, this is a well known and useful length scale for estimating the relative scaling of self-steepening which we adopt here for convenience. Note that the mechanism presented here is not the only possibility for developing shock fronts. A highly nonlinear medium in presence of weak normal dispersion, for example, could also lead to shock formation even though no dispersive nonlinearity is present \cite{Trillo:PRA14}.

The nonlinear dispersion $\tau_{NL}$ plays a key role in the pulse dynamics of both the temporal and spectral properties. We first address the impact of SS on temporal shape and delay. A subtle point that must be addressed is the dual role of $\beta_2$. First, it has been shown in earlier work that $\beta_2$ dissipates shock fronts \cite{fiberSS}. Second, and separate to this point, we showed above that $\beta_2$ is intrinsically linked to the large magnitude of $\tau_{NL}$. As a result, one cannot ignore $\beta_2$ for a SS effect arising from a strong modulation of the group index $n_g$ and consequently observing a tilted line-shape is unlikely for SS derived from this method. While the temporal shape is not modulated in this case, the temporal arrival time is affected as we show below.

\begin{figure}[h]
\centerline{\includegraphics[width=9cm]{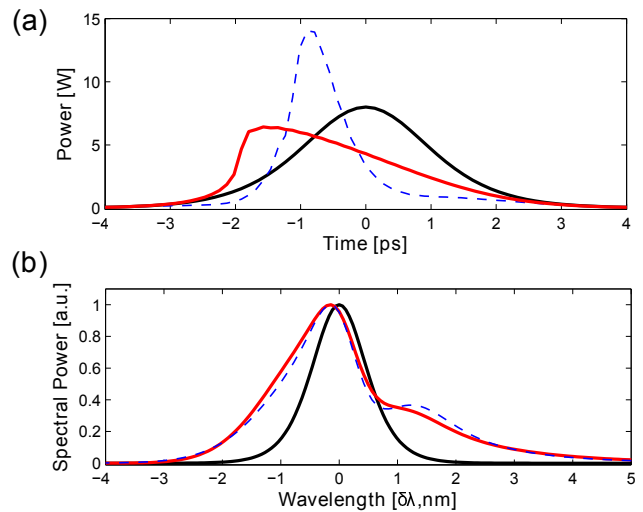}}
\caption{Temporal and spectral pulse properties of \textit{anomalous} self-steepening in PhC waveguides. (a) Temporal pulse shapes. (b) Spectral pulse shapes. Notice the behavior is opposite to unstructured media exhibiting normal self-steepening. Traces are shown for propagation of 200~$\mu$m$~\approx~2.5 L_{NL}$. Legend: Kerr and SS only (thick red), full Eqn. \ref{eq:GNLSE} (dashed blue), and input pulse (black).}
\label{fig:negativeSS} 
\end{figure}

\figs{fig:negativeSS}(a) shows the temporal profile obtained for the point indicated by the blue dot in \fig{Fig:SS_PhCWG}(a-c) after a propagation distance of 200~$\mu m$. The input pulse (black line) has a peak power of 8~W ($L_{NL}$=80 $\mu$m), hence the pulse has completed about two and a half nonlinear lengths $L_{NL}$.  The thick-red line shows the case where we include only the Kerr and SS contributions to Eqn. \ref{eq:GNLSE}. That is, we include the $\beta_2$ contribution to $\tau_{NL}$ but neglect the temporal dispersion term $\partial_{tt}$. Notice that since $s$ is negative here the pulse peak tilts \textit{forwards} in time, which is opposite to earlier studies with a positive self-steepening term. Moreover, with the large $s$ value here, a steep temporal leading edge is already clearly visible after a propagation of a just few $L_{NL}$. Simulations with all the terms in Eqn. \ref{eq:GNLSE} are shown as the dashed blue line. In contrast to the ideal dispersion-less case (thick red) where the temporal trace exhibits a clear abrupt leading edge characteristic of shock formation, the shock front is less pronounced for the same propagation length and the pulse tends to preserve its initial shape (soliton effect). 

A separate temporal effect in the presence of self-steepening is a shift in pulse arrival time. Assuming the pulse duration $T_0>\tau_{NL}$ and a moderate soliton number, the dispersive nonlinearity acts as a small perturbation that slightly modifies the group velocity $\Delta T=\frac{z}{L_{NL}}\tau_{NL}$ \cite{anderson1983,PhysRevE.57.4751,chen2010soliton}. Here with the negative $\tau_{NL}$ value, the pulses are expected to advance in time, once again in contrast to the delay of earlier observations. This effect is clearly visible by the temporal advance of the two traces examined in \fig{fig:negativeSS}. One of the most challenging aspects of observing self-steepening in real PhCWGs is the strong resemblance of SS and free-carrier dispersion (FCD) in the time domain. We investigate this in detail below. 

\figs{fig:negativeSS}(b) shows typical pulse spectra for the negative $s$ values characteristic of PhCWGs. The pulse spectrum is clearly rendered asymmetric in the presence of self-steepening. The blue-shifted peaks become more intense, while the red components are notably broader compared to the symmetric broadening characteristic of SPM.  The role of $\beta_2$ is less pronounced.


\section{Self-steepening in semiconductors}
In our analysis thus far we have ignored several effects present in practical systems. The role of linear loss has been previously studied and in the limit $\alpha z_s>$1 shock waves are unobservable altogether \cite{anderson1983}. For the waveguide experiments referenced above the linear loss is given by $\alpha$~=~15 dB/cm~(at~$n_g$~=~15). This yields $\alpha z_s$~=~0.11, indicating these waveguides can in principle support shocks. 

However, in semiconductor media we must also consider nonlinear absorption of multiple photons across the electronic bandgap. For a typical wavelength of 1550 nm (photon energy of 0.8 eV), silicon ($E_g$~=~1.1 eV) is restricted by two-photon absorption (TPA) and the wide-gap material GaInP ($E_g$~=~1.9 eV) is limited by three-photon absorption (3PA). While the optical properties of silicon have been widely studied, only over the past years we have investigated the $\chi^{(3)}$ properties of GaInP and established this material as a viable platform for nonlinear optics at 1.5~$\mu$m~\cite{Husko_SciRep13,NatPhot_Colman10,huskoOptExp2009,combrie2009}. In simplest terms, nonlinear absorption damp the dynamics similar to linear absorption. An order of magnitude estimate shows loss due to TPA (3PA) requires $\alpha_2 I z_s<1$ ($\alpha_3 I^2 z_s<1$), with the intensity $I=\frac{P_o}{A_{eff}}$, to observe a shock. Physically these ratios compare the strength of the nonlinear loss to the self-steepening term.

For the TPA case (e.g. Silicon, $\alpha_{2}$=1 GW/cm \cite{bristow2007}) the ratio $\alpha_2 I z_s=\frac{0.43}{|s|}~\frac{\alpha_2 }{k_o n_2}\approx 1.8$ indicating SS-induced shocks are not generally accessible in this material, even for this large $s$=~-0.1. Here we have defined a new nonlinear figure-of-merit (FOM) for self-steepening and TPA. Its form is noticeably similar to the well-known version for Kerr-TPA switching $(\frac{\alpha_2}{k_on_2})$ \cite{mizrahi1989} with the additional term for SS. Notice this ratio is independent of power for TPA and therefore SS will always be much weaker than TPA. Since TPA dominates the SS effect, we focus mainly on the 3PA system in the following analysis. 

The 3PA material (e.g.~GaInP, $\alpha_{3}$=0.013~GW$^2$/cm$^3$ \cite{wherrett1984}) is much more favorable, yielding $\alpha_3 I^2 z_s=\frac{0.43}{|s|} \frac{\alpha_3 I}{k_o n_2}~\leq$~1 which is satisfied for intensities up to about $\approx$50~GW/cm$^2$. This threshold is intensity-dependent due to the different nonlinear orders of $\chi^{(3)}$ Kerr and $\chi^{(5)}$ 3PA. Including the slow-light enhancements $\alpha_3\propto n_g^3$ and $n_2 \propto n_g^2$ this relation becomes $\frac{0.43}{|s|} \frac{\alpha_3 I}{k_o n_2} \left(\frac{n_g}{n_o}\right)\leq$~1 and is satisfied up to $\approx$10~GW/cm$^2$ for the conditions here. In contrast, the TPA-limited SS case does not scale with slow-light. Note these estimates are only indicative and do not correspond to strict thresholds. For larger powers, free-carrier absorption would play an important role until the peak power falls below its threshold. Now that we have established the role of nonlinear loss, we describe the free-carrier effects.

Free-carriers generated via these nonlinear absorption mechanisms have an equally significant impact in the pulse dynamics through both dispersive (FCD, $n_{FC}$) and absorptive (FCA, $\sigma$) contributions. Importantly, the physical signatures of anomalous self-steepening strongly resemble those of FCD, especially in the time domain. Namely, like anomalous SS, the pulse also advances as a function of input power due to FCD combined with anomalous GVD as our recent experiments show \cite{Husko_SciRep13,blancoFCD2014}. 

A broader GNLSE including all of these effects is:
\begin{multline}
\frac{\partial A}{\partial z} =-\frac{\alpha}{2}A - i\frac{\beta_2}{2} \frac{\partial^2A}{\partial t^2} + \frac{\beta_3}{6} \frac{\partial^3A}{\partial t^3}
 +(ik_o n_{FC}-\frac{\sigma}{2})N_cA \\
 + (i\gamma_{eff} - \gamma_1 \frac{\partial}{\partial t}-\frac{\alpha_{2eff}}{2})|A|^2A -\frac{\alpha_{3eff}}{2}|A|^4A.
\label{eqn:fullNLSE}
\end{multline}

\begin{figure}[b]
\centerline{\includegraphics[width=9cm]{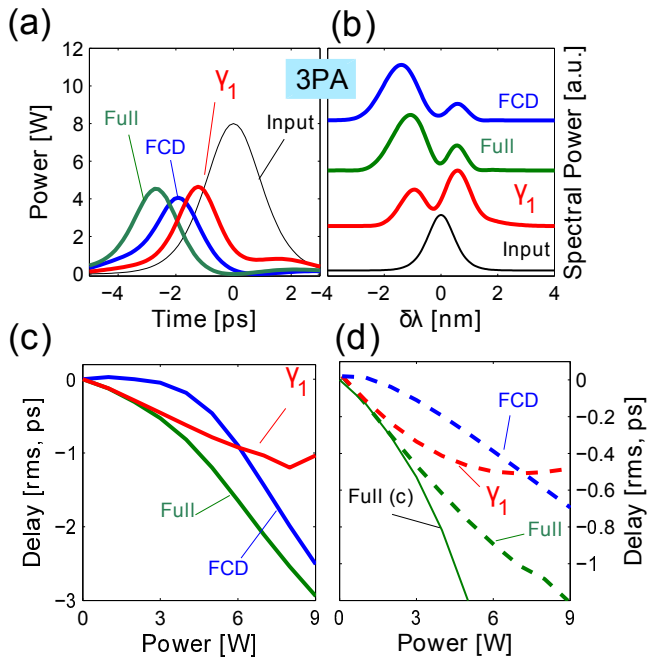}}
\caption{Temporal and spectral behavior of anomalous self-steepening in after a propagation distance of 500~$\mu m$ in realistic PhCWGs. (a) Temporal traces for 3PA-limited materials: full (green), FCD alone (blue), self-steepening alone (red). The black lines correspond to the input pulse. (b) Spectra for 3PA-limited materials. (c)-(d) Pulse temporal advance as a function of power for (c) 3PA-limited (solid) and (d) TPA-limited (dashed) systems. Green curves correspond to the full model in Eqn. \ref{eqn:fullNLSE}. Blue lines are the case where we neglect SS, whereas red lines indicate when FCD is neglected. We include the full 3PA result in (d) to compare the relative scales of the two systems.
}
\label{fig:temporalSpectral} 
\end{figure}

\noindent For the modeling below, we take the slow-light scaled values of the bulk parameters as described in prior literature \cite{huskoOptExp2009,blancoFCD2014}. The parameters are  $\alpha_{3eff}=\frac{\alpha_3}{A_{5eff}^2}\left(\frac{n_g}{n_0}\right)^3$=~18~m$^{-1}$.W$^{-2}$, $n_{FC,eff}$=$n_{FC}\left(\frac{n_g}{n_0}\right)$=~-1.8 $\times10^{-26}$m$^3$, $\sigma_{eff}=\sigma\left(\frac{n_g}{n_0}\right)$=~1.3 $\times10^{-20}$m$^2$. The 3PA area is $A_{5eff}$\cite{huskoOptExp2009}. We have also included the third-order dispersion (TOD, $\beta_3=\frac{1}{c}\frac{d^2n_g}{d\omega^2}$~=~+0.7 ps$^3$/mm) for completeness as this cannot be ignored in the real system. Regarding one case briefly presented below where the material is limited by TPA (silicon), we take $\alpha_{2eff}=\frac{\alpha_2}{A_{eff}}\left(\frac{n_g}{n_{0,Si}}\right)^2$~=~570~(W.m)$^{-1}$. It is worthwhile to point out that these parameters correspond to actual experimental parameters and therefore are immediately realizable in current systems.

\figs{fig:temporalSpectral}(a) shows the pulse temporal shift due to the competing self-steepening and FCD effects in the 3PA limited material ($\alpha_2=0$) at a peak power of 8~W. When considered in isolation the FCD-GVD curve (blue, no SS) and SS-only curve (red, no FCD) each would contribute a few picoseconds of delay. Moreover they have a relatively similar temporal shape. Note these effects do not add linearly, but rather compete for power and interact dynamically to yield the full result (green). 

Importantly, the spectral features of FCD and SS are distinct. \figs{fig:temporalSpectral}(b) shows the spectral properties of pulses propagating in the 3PA system. Considering only the self-steepening effect (red curve) results in a minor modulation to the symmetric shape expected from SPM-only and does not shift the spectral center-of-mass. In contrast, the FCD induces a clear blue shift in the pulse center-of-mass \cite{Husko_SciRep13}. When we consider these effects simultaneously (green curve), the FCD is clearly the dominant contribution. Thus these effects are more easily distinguished in the spectral domain.

The nonlinear scaling laws of SS and the FCD-GVD temporal shift are also notably different. Notice $\tau_{NL}~\propto~P_o$ whereas FCD$^{(2)}\propto~P_o^2$ (TPA) or FCD$^{(3)}\propto~P_o^3$ (3PA), where we have written FCD$^{(m)}$, with $m$ indicating the order of nonlinear absorption generating the free-carriers. Figures \ref{fig:temporalSpectral}(c) and (d) report the scaling of SS and FCD as a function of power for (c) 3PA and (d) TPA-limited materials. We observe SS (red curve) scales much more slowly compared to FCD (blue curve). The full simulation (green dashed) more closely follows SS at low powers and the FCD trend at higher power. For the TPA case shown in (d) the SS induced delay is noticeably smaller due to stronger nonlinear TPA loss which caps the peak power more than the 3PA case. In (d) we also show the 3PA curve for comparison, highlighting the greater temporal shift in this case.

The striking similarity of the temporal advance and pulse shapes from both SS and FCD-GVD are highlighted in recent experimental results. The GaInP waveguides in both cases are similar to that in \fig{Fig:SS_PhCWG}(c), with relatively small values of $\tau_{NL}$ compared to that highlighted in this work. In a ThPA system the temporal advances are attributed to FCD-GVD supported with numerical modelling \cite{Husko_SciRep13}, however with no SS term. Similar results were shown in Refs.\cite{blancoFCD2014,blanco2014observation} in silicon. In contrast, a pulse advance attributed to SS only was reported in Ref. \cite{Raineri_PRB13}. Critically that report did not include the important contribution of FCD-GVD term, but rather attributed it to SS alone. As we have shown here, the FCD-GVD advance plays an equally important role as SS and cannot be ignored. 

The true physical situation is likely a combination of these effects though the exact scaling would be challenging due to soliton dynamics modulating the peak power and requires careful experiments and numerical modeling to discern the effects. Nonetheless, one could use the moments method in Ref. \cite{lefrancois2015} to estimate the magnitude of the temporal shift from FCD-GVD or the SS time shift equation given above for order-of-magnitude values.

One last important aspect to consider is the relative impact of the instantaneous SS and non-local FCD effects on pulse delay during propagation. The SS induced delay depends directly on the instantaneous peak power and hence will eventually decrease due to linear and nonlinear loss. In contrast, the effect of FCD on the group velocity persists even if the pulse peak power decreases. Physically this results from the spectral FCD frequency blue-shift and accompanying change in the group-velocity experienced by the pulse \cite{lefrancois2015}. 

\begin{figure}[h]
\centerline{\includegraphics[width=6cm]{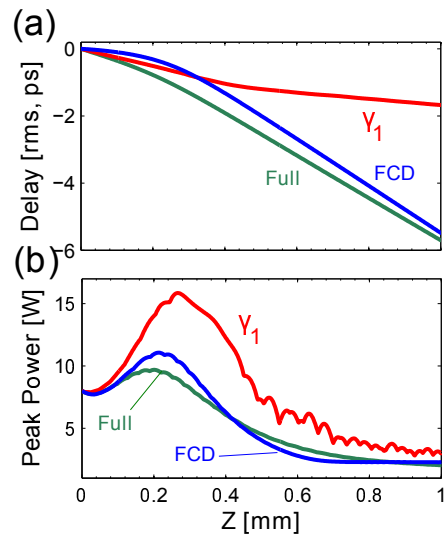}}
\caption{Role of instantaneous SS and non-local FCD on pulse delay. (a) Contribution of SS alone (red) and FCD alone (blue) to the total pulse delay (green) for propagation lengthts up to 1 mm. (b) Peak power evolution corresponding to the cases presented in the upper panel. Note the low overall peak power after 500 $\mu$m of propagation. Legend: full model (green), no free carrier effects (red), no SS (blue).
}
\label{fig:Delay_Z} 
\end{figure}

Finally we show the evolution of the nonlinear dynamics of the pulse propagating down the waveguide. \figs{fig:Delay_Z}(a) shows the change in pulse delay along the waveguide length for the two mechanisms in the 3PA system at a peak power of 8~W. The change in delay due to the SS effect (red, FCD=0 and FCA=0) is strongest in the first 200 $\mu$m and then tapers off due to loss. The FCD effect (blue, $\gamma_1$=0) continues to experience a change in delay right until the end of the propagation with the full curve (green) more closely resembling this case. \figs{fig:Delay_Z}(b) shows the corresponding peak power evolution which gives insight into the dominant nonlinear mechanism as the pulse evolves. We attribute the initial increase in peak power to soliton compression.
\vspace{-1 cm}

\section{Conclusion}
In this Letter, we investigated the nonlinear self-steepening effect in photonic crystal waveguides. Our first principles derivation in the nanostructured periodic waveguides revealed a self-steepening term two orders of magnitude larger than typical systems. Importantly the self steepening coefficient $\tau_{NL}$ is determined by the geometric parameters of the waveguide, offering a large tuning range of both positive and negative values. The origin of this giant $\tau_{NL}$ is the strong dispersion of PhCWGs counterbalanced by a modal area contribution. We considered the role of higher-order effects in practical systems with new figures of merit, concluding that the nonlinear loss quenches the dynamics. We showed that the principal physical signature of the \textit{anomalous} SS effect is a temporal forward tilt and pulse advance in contrast to the delay observed in normal SS media. In the semiconductor waveguides these effects compete with FCD which also advance the pulse, whereas the spectral signatures of self-steepening and FCD are distinct. We suggest future experiments exploring the full range of these dynamics be undertaken to reveal the full dynamics of this giant tunable self-steepening mechanism, especially in the supercontinuum regime.
\\

\noindent \textbf{Acknowledgements}\\ 
This work was supported by the VKR fundet through the centre of excellence NATEC and the Australian Research Council (ARC) Discovery Early Career Researcher Award (DECRA DE120102069).


%

\end{document}